\documentclass[twocolumn,
prd, 
nofootinbib, tightenlines,
showpacs, 
preprintnumbers, amsmath, amssymb]
              {revtex4}
\usepackage{graphicx} 

\begin{document}
\title{ Unification and mass spectrum in the minimal $B - L$ model}


\date{\it May 2011}
\author{ R. J. Hern\'andez-Pinto} \email{rhernand@fis.cinvestav.mx}
\affiliation{ Departamento de 
F\'isica, Centro de Investigaci\'on y de Estudios Avanzados del I.P.N., 
Apdo. Post. 14-740, 07000 M\'exico D.F., M\'exico. }

\author{ A. P\'erez-Lorenzana}
\email{aplorenz@fis.cinvestav.mx} \affiliation{ Departamento de 
F\'isica, Centro de Investigaci\'on y de Estudios Avanzados del I.P.N., 
Apdo. Post. 14-740, 07000 M\'exico D.F., M\'exico. }

\begin{abstract}
Gauging $B-L$ symmetry provides a simple realization of the seesaw mechanism in a naturally anomaly free extension to the MSSM gauge group, 
$SU(3)_c\times SU(2)_L\times U(1)_Y \times U(1)_{B-L}$. 
However, as we discuss in here, it turns out that the simplest $B-L$ extension of the MSSM may change some of the conceptions about the path for gauge unification as well as to
affect the predicted spectrum of the supersymmetric particles at low energy. For instance, the coupling $g_{B-L}$ ended up to be smallest of all of them and the lightest gaugino is the one related to $B-L$. RGE also help to understand the spontaneous breaking of the $U(1)_{B-L}$ and the vacuum expectation value of the sneutrino at low energies, which occurs in this type of models.
\end{abstract}

\maketitle
\section{Introduction}
The Stantard Model (SM) of particles physics have been very successful on explaining with an impressive accuracy several experimental results. Collider phenomenology support it in almost all the processes measured. Nevertheless, the SM cannot explain completely the neutrino phenomenology~\cite{Neu}, the dark matter content in the universe~\cite{DM}, and recently  it has a tension in the anomalous like-sign dimuon charge asymmetry published by the D$\O$ experiment at Fermilab~\cite{Assym}. All these discrepancies have opened a new window in order to build theories on top of the SM. Theories such as Grand Unified Theories, Supersymmetric Theories, etc., could explain some experimental facts that, by itself, the SM cannot.

The fact that the neutrino has zero mass in the SM is not consistent with the neutrino oscillation experiments~\cite{Osc}. This result can be easily explained by adding a neutrino mass term to the SM lagrangian. In order to write a mass term for neutrinos, it is needed to deal with the inclusion a right handed neutrino field $\nu_R$, which is not considered in the SM. Then, it would be possible to write the Dirac and Majorana masses,
\begin{equation}\label{nMSM}
\delta \mathcal{L}=i\bar{\nu}_R\partial_{\mu}\gamma^{\mu}\nu_R-h\sigma\bar{\nu}^c_R\nu_R-h'\bar{L}\tilde{H}\nu_R.
\end{equation}
This extension to the SM has been considered to give some explanation to dark matter, barion asymmetry, inflation, neutrino mass and oscillations~\cite{Sha205, Boy106, Boy206, Lai06, Lai07, Dod94, Shi99, Dol02, Aba01, Kus97, Kuz85}. However, the theory described by Eq.~\eqref{nMSM} breaks explicitly the $B-L$ global symmetry which underlay in the SM. The connection between the neutrino mass term and the Higgs mechanism suggest an extended gauge group $SU(2)_L\times U(1)_Y\times U(1)_{B-L}$ as a natural symmetry to extends the electroweak group. This gauge group contain an extra gauge coupling and a gauge boson which would be present in the already measured cross sections. Several studies has been implemented in $B-L$ theories, including right handed fields~\cite{a}, supersymmetric theories with broken R-parity~\cite{b}, collider phenomenology~\cite{c}, etc., and all of them have interesting features such as the breaking of the $B-L$ gauge group radiatively, or the possible signals that could be seen at LHC, among others.

The parameters of a model take different values depending on the energy scales where they are measured, and its behavior is controlled by the renormalization group equations (RGE). And, it is important to consider the extra parameters of a model, because they could give some indications of new physics, but in order to give some predictions it is necessary to know its values at the energies tested nowadays. On the other hand, it is important to know, by experimental facts, the initial conditions of these differential equations if it is pretended to give their values at some desired scale. The RGE formalism has been used in the SM and it has showed that the desired unification is not achieved in it~\cite{GaugUnif}, but in its minimal supersymmetric extension, so called MSSM. The MSSM is an attractive theory; quadratic divergences and the hierarchy problem are absent; the unification of the gauge couplings and the spontaneous symmetry breaking of $SU(2)_L\times U(1)_Y$ is naturally achieved by the RGE, and it also have a dark matter candidate, when the $R-$parity is conserved. If $U(1)_{B-L}$ is included $R-$parity does not need to be imposed because the operators,
\begin{eqnarray}
W &\supset& \lambda ^{ijk} L_iL_j\bar{e}_k +\lambda'^{ijk} L_i Q_j \bar{d}_k +\mu'^i L_i H_u \nonumber\\
&&+\lambda''^{ijk} \bar{u}_i\bar{d}_j\bar{d}_k
\end{eqnarray}
included in the superpotential cannot be written due to the violation of the $B-L$ symmetry and therefore this extension have naturally a dark matter candidate~\cite{BLRParity}.

In this paper we will focus on a supersymmetric model which could, in principle, give an explanation to neutrino masses. In order to explain the extra terms in the lagrangian we gauge the fields under $\mathcal{G}=SU(3)_c\times SU(2)_L \times U(1)_Y \times U(1)_{B-L}$; we are interested in low energy phenomenology thus we calculate and solve the renormalization group equations in order to get low energy values for the masses and couplings. 

This article is organized as follow, in Section \ref{Model} we write the extra superfields, and the transformation laws that they obey under the gauge group $\mathcal{G}$, we write also the additional superpotential, the soft breaking lagrangian and the corresponding Kh\"aler potential for the superfields, the gauge lagrangian and the scalar potential for the extra Higgses; in Section \ref{couplings} we calculate the beta function for the gauge couplings and we present the solution for a unified scenario; in Section \ref{ZBL} we analyze the applicable constraints given by $Z'$ searches that could restrict our parameter space; in Section \ref{sparticles} we present the beta functions for the mass parameters, Yukawa couplings and the anomalous dimensions and we solve them for two scenarios; in Section \ref{vev} we analyze the low energy behavior of vacuum expectation value of the sneutrino, and then, we present our conclusions.

\section{The model}\label{Model}
We present a model based in the supersymmetric extension of the gauge group $\mathcal{G}$. 
The matter superfields includes all of those of the MSSM, and the right handed neutrino superfield, $N$. 
The supersymmetric extension of Eq.~\eqref{nMSM} requires two extra Higgses, $\sigma_1$ and $\sigma_2$. 
Then, under $\mathcal{G}$ these superfields transform as,
\begin{eqnarray}
\bar{\hat{N}}&\sim&\left(\mbox{\bf{1},\bf{1}},0,1\right)\\
\hat{\sigma}_1&\sim&\left(\mbox{\bf{1},\bf{1}},0,-2\right) \\
\hat{\sigma}_2&\sim&\left(\mbox{\bf{1},\bf{1}},0,2\right)
\end{eqnarray}

In order to describe the pieces of the model, we define the contributions that we are going to add to the MSSM by $\Delta \omega= \omega_{\mathcal{G}}-\omega_{MSSM}$, where $\omega$ will be the superpotential, the soft symmetry breaking lagrangian, and the K\"ahler potential . We start by defining the superpotential; the most general superpotential that can be written is,
\begin{eqnarray}\label{superpotential}
\Delta W&=& 
\bar{\hat{N}}\mbox{\bf{Y}}_N^D\hat{L}\hat{H}_u+\hat{N}\mbox{\bf{Y}}_N^M\hat{N}\hat{\sigma}_1+\mu^{\prime}\hat{\sigma}_1\hat{\sigma}_2.
\end{eqnarray}
where $\mbox{\bf{Y}}_N^M$ and $\mbox{\bf{Y}}_N^D$, are the Yukawa matrices corresponding to the Majorana and Dirac's terms respectively. 
As we have pointed out in the introduction we do not need to impose $R-$parity, $U(1)_{B-L}$ avoid us to write the operators that provoke the instabilities of the proton and moreover we have naturally a dark matter candidate, the lightest supersymmetric particle (LSP).

Applying the general prescription for the soft supersymmetric breaking terms, we add the following terms to the Lagrangian,
\begin{eqnarray}
-\Delta\mathcal{L}_{SB}&=&
\tilde{\bar{N}}\mbox{\bf{h}}_N^D\tilde{L} H_u 
+\tilde{N}^c\mbox{\bf{h}}_N^M\tilde{N} \sigma_1 
+\tilde{N}^{\dagger}{\bf m^2_N}\tilde{N}
 \nonumber\\
&& +m^2_{\sigma_1} \sigma^{\dagger}_1 \sigma_1+m^2_{\sigma_2}\sigma^{\dagger}_2\sigma_2+B^{\prime}\sigma_1\sigma_2
\nonumber\\
&&
+ \frac{1}{2}M_{B-L}\tilde{Z}_{B-L}\tilde{Z}_{B-L} 
\end{eqnarray}
where the tilde fields are the superpartners of the fields involved in the superpotential, and $\tilde g$, $\tilde W$, $\tilde B$ and $\tilde Z_{B-L}$ are the gauginos of the theory.

For completeness, the additional K\"ahler potential is given by,
\begin{eqnarray}
\Delta K= \hat{N}^{\dagger} e^{2V} \hat{N} + \hat{\sigma}_1^{\dagger} e^{2V} \hat{\sigma}_1 +\hat{\sigma}_2^{\dagger} e^{2V} \hat{\sigma}_2
\end{eqnarray}
and the gauge lagrangian is
\begin{eqnarray}
\mathcal{L}= \frac{1}{4} \int d\theta^2 W^{\alpha}_{(B-L)}W_{\alpha (B-L)} + h.c. 
\end{eqnarray}
where
\begin{eqnarray}
\hspace{-6mm} W^{\alpha}_{(B-L)}W_{\alpha(B-L)}\vert_{\theta\theta}= -2i \tilde{Z}_{B-L}\sigma^{\mu}\partial_{\mu} \bar{\tilde{Z}}_{B-L}+D^2 \nonumber \\
 -\frac{1}{2}A_{\mu\nu}A^{\mu\nu} -\frac{i}{4}\tilde{A}_{\mu\nu}A^{\mu\nu},
\end{eqnarray}
in the WZ gauge; and $A^{\mu\nu}=\partial^{\mu}Z^{\nu}_{(B-L)}-\partial^{\nu}Z^{\mu}_{(B-L)}$, the dual is $\tilde{A}^{\mu\nu}=\epsilon^{\mu\nu\alpha\beta}A_{\alpha\beta}$; and 
$\tilde Z_{B-L}$ is the superpartner of the $Z_{B-L}$ gauge boson associated to $U(1)_{B-L}$ gauge group.

Finally, we need the scalar potential for the Higgs fields which have the form,
\begin{eqnarray}
V(H_u,H_d,\sigma_1,\sigma_2)&=&m_{Hu}^{\prime 2}|H_u|^2+m_{Hd}^{\prime 2}|H_d|^2\nonumber\\
&&\hspace{-20mm}-(BH_uH_d+c.c.)+\frac{1}{8}g^{\prime}(|H_u|^2-|H_d|^2)^2\nonumber\\
&&\hspace{-20mm}+m_{\sigma_1}^{\prime 2}|\sigma_1|^2+m_{\sigma_2}^{\prime 2}|\sigma_2|^2\nonumber\\
&&\hspace{-20mm}-(B^{\prime}\sigma_1\sigma_2+c.c.)+\frac{1}{8}g_{B-L}(|\sigma_1|^2-|\sigma_2|^2)^2,\nonumber\\
\end{eqnarray}
where $m_i^2=m_i^{\prime 2}+\mu^2$ for $i=H_u, H_d$, and $m_j^2=m_j^{\prime 2}+\mu^{\prime 2}$ for $j=\sigma_1, \sigma_2$. 


\section{One loop unification of gauge couplings}\label{couplings}
In order to get low energy values of the model, we use the RGE formalism to calculate the beta functions of masses, couplings and the anomalous dimension parameters. We are interested to find the low energy value of $\alpha_{B-L}$, therefore we calculate the beta function of it, and we will solve it by implying a unification at the GUT scale.
For a general superpotential and soft breaking terms, the $\beta-$functions can be calculated by using the representations of the superfields~\cite{Mar94}. In the following, we will use the
notation: $\displaystyle \beta_f=16\pi^2 (df/dt)$, where $t=\ln(Q/Q_0)$, $Q$ is the renormalization scale, and $Q_0$ is the reference scale.

The one loop beta functions for the gauge coupling constants are given by,
\begin{equation}
\beta _{g_i}=c_ig^3_i,\quad\mbox{where}\quad i=1,2,3,B-L
\end{equation}
where $c_i$ corresponds to the embedding factors. The unification of the gauge couplings $\alpha_i$ occurs when the values of all match into a common value, $\alpha=g^2/4\pi$ where $\alpha$ is the corresponding gauge coupling of the unifying group, $G$. Due to $G \supset\mathcal{G}$, we need to normalize the generators in order to keep the relation $\alpha_i=c_i\alpha$ at the unification scale. It is important to notice that one can calculate the $c_i$ factors regardless on our ignorance the actual group that will realize unification, provided it keeps family universality on gauge interactions. The embedding factors are, in general, rational numbers satisfying,
\begin{equation}
c_i\equiv\frac{\mbox{Tr } T^2}{\mbox{Tr } T^2_i}
\end{equation}
where {\it T} is a generator of the subgroup $G_i$ normalized over a representation $R$ of $G$, and $T_i$ is the same generator but normalized over the representations of
$G_i$ embedded into $R$~\cite{Per00}. Computing the embedding factors we have,
\begin{equation}
(c_1,c_2,c_3,c_{B-L})=(3/5,1,1,3/8),
\end{equation} 
where $c_{1,2,3}$ are those well known for standard path of unification. It is worth noticing that $c_{B-L} \neq c_1$.

The general solution for $g_i$ can be expressed in terms of $\alpha_i\equiv g^2_i/4\pi$, and the solution is~\cite{Per98},
\begin{equation}\label{Callan}
\alpha^{-1}_i(m)=c_i\left[\alpha^{-1}_i(m_Z) +(2\pi)^{-1}b_i\ln \left(\frac{m}{m_Z}  \right)   \right].
\end{equation}

It is left to calculate the $b_i$ constants, we use the general formulae for a supersymmetric theory~\cite{Bailin},
\begin{equation}
b=3C_1(G)-\sum_R C_2(R)
\end{equation}
where $C_1(G)$ is the quadratic Casimir invariant of the $G$ group, and $C_2(R)$ the Dynkin index of the $R$ representation.
Therefore, these constants have the values,
\begin{equation}
(b_1,b_2,b_3,b_{B-L})=(-11,-1,3,-24).
\end{equation}

In order to solve the equation \eqref{Callan}, we need to impose initial conditions; we use for 
$\alpha^{-1}_1(m_Z) \approx 98\mbox{.}33 , \alpha^{-1}_2(m_Z) \approx 29\mbox{.}57  , \alpha^{-1}_3(m_Z) \approx 8\mbox{.}4 $~\cite{PDG}.
However, we have none information about $\alpha_{B-L}^{-1}(m_Z)$. For this purpose, we have supposed that all $\alpha_i^{-1}$ will unify 
at the $M_{GUT}$ scale, which corresponds to $M_{GUT}\sim 2.5 \times 10^{16}$  GeV. 
With this consideration, its possible to plot the corresponding running provided by Eq.~\eqref{Callan}. 

With the solution showed in FIG. 1, we can calculate $\alpha^{-1}_{B-L}(m_Z)\approx 191\mbox{.}1$ and, $g_{B-L}(m_Z)\approx 0\mbox{.}2565$, which
is consistent with previous findings in Ref.~\cite{Her109}. 
\begin{figure}[t!]
\begin{center}
\includegraphics[scale=.44]{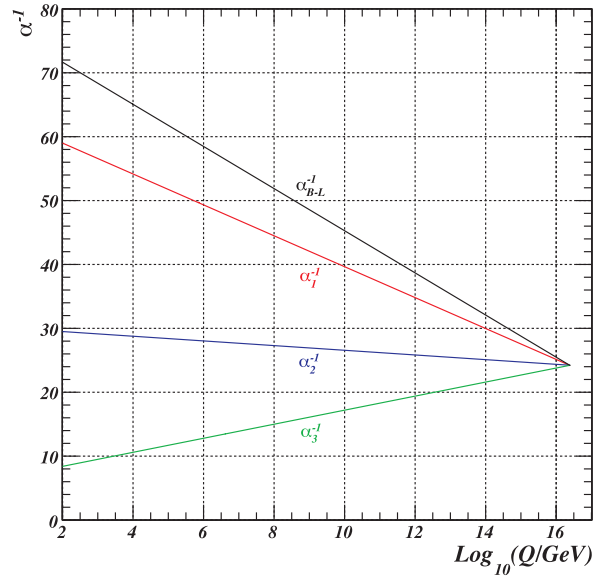}
\caption{Running of the coupling constants for $\mathcal{G}-$SUSY model. In the running we have assumed that there is a unification of all the couplings at the GUT scale.}
\end{center}
\end{figure} 

Now, we need to look back to some experimental results because the inclusion of $Z_{B-L}$ lead to contributions to all experiments in which the $Z^0$ gauge boson is participating.


\section{Limits on the $Z_{B-L}$}\label{ZBL}

The net effect of extending the SM gauge group with a $U(1)$ is that the theory will have, in general, an extra $Z'$ gauge boson. This gauge boson have to mix with the $Z^0$ of the SM and its effects will appear in all the tested processes in which the $Z^0$ is present. 

In our case, this two gauge bosons do not mix; by taking a look to the K\"ahler potential we have,
\begin{eqnarray}
\mathcal{L} &=& \left[ D^2H_u\right]^{\dagger}  H_u + \left[ D^2H_d\right]^{\dagger}  H_d \nonumber \\
&&+ \left[ D^2 \sigma_1\right]^{\dagger} \sigma_1 + \left[ D^2 \sigma_2\right]^{\dagger}  \sigma_2  + \left[ D^2 \tilde N\right]^{\dagger} \tilde N; 
\end{eqnarray}
where $D^2=D_{\mu}D^{\mu}$, and the gauge covariant derivatives are,
\begin{eqnarray}
D_{\mu} H_i &=& \left ( \partial_{\mu}  +ig W_{\mu} +ig' YB_{\mu} \right)H_i \quad i=u,d\\
D_{\mu} \Psi_\ell &=& \left ( \partial_{\mu}  +ig_{B-L} Y_{B-L}Z_{B-L} \right)\Psi_\ell
\end{eqnarray}
and $Y=+1/2$ for $H_u$ and $Y=-1/2$ for $H_d$, and we have identified $\Psi_1 \equiv \sigma_1 $, $\Psi_2 \equiv \sigma_2 $ and $\Psi_3 \equiv \tilde N $ and its respective $B-L$ hypercharges has been specified in Eq. (3-5). We have included the sneutrino in the discussion because several authors~\cite{nuvev} have shown that in some scenarios the sneutrino field could acquire a vacuum expectation value, and if it is so then it will contribute to the mass of the $Z_{B-L}$ gauge boson. So in the most general case, when these fields acquire a vacuum expectation value the masses of the gauge fields will be,
\begin{eqnarray}
m_Z^2&=&\frac{1}{4}\left( g^2+g'^2\right)v^2 \\
m_W &=& m_Z c_{\theta_W} \\
m_{Z_{B-L}}^2 &=& g_{B-L}^2 v_{B-L}^2
\end{eqnarray}
where $v^2 =v_u^2 +v_d^2\approx (246 \mbox{ GeV})^2 $ and $v_{B-L}^2 = 4(v_1^2+v_2^2)+v_{\tilde N}^2$, and from which we have no information about its scale; the electroweak angle $\theta_W$ is introduced in order to rotate the fields into a eigenstate mass basis, $W^0, Z^0, W^{\pm}$; this feature it is not needed for the $B-L$ sector because the SM-gauge fields are not gauged under $U(1)_{B-L}$ and so, $B-L$ is already in mass eigenstate. High order operators could mix this two gauge bosons; operators in the K\"ahler potential as
\begin{eqnarray}
K \supset \sum_{i=u,d}\sum_{j=1}^3 \frac{c_{ij}}{M^2}(\hat{H}_i^{\dagger}e^{2V}\hat{H}_i)(\hat{\Psi}_j e^{2V}\hat{\Psi}_j)
\end{eqnarray} 
where $c_{ij}$ are order 1 coefficients. If $c_{ij}\neq 0$ then they will definitively provoke a mixing between $Z^0 - Z_{B-L}$, although they will contribute to the precision measurements, the mass of the SM gauge boson, the invisible decay of the $Z^0$, etc., and they could be constrained strongly by all these tests.  Moreover, these operators are suppressed by a mass scale $M$ to the second power and its scale could be, in the best scenario, of the order of TeV where the new physics would appear. 
Due to the fact that we are not considering higher order operators, $c_{ij}=0$, we can treat both gauge bosons as mass eigenstates, and
the fact that using RGE we could get a prediction for $g_{B-L}$ has a direct impact on the mass of the associated $U(1)_{B-L}$ gauge boson. The constraint obtained for a $B-L$ gauge boson is writen as~\cite{Carena}, $
M_{Z_{B-L}} \geq g_{B-L} \times \mbox{6 TeV},
$
therefore, in our model is translated to,
 \begin{eqnarray}
 M_{Z_{B-L}} \geq \mbox{1.5 TeV},
 \end{eqnarray}
and within this value we are also in agreement with a recent analysis~\cite{Aguila}. Moreover, it also means that the $v_{B-L}$, which in general would contain contributions for the $\sigma$ scalars and the sneutrino, has to be of the order of a few TeV, which is still sizable for the LHC.


\section{Superparticles mass spectrum}\label{sparticles}
\begin{figure}[h!]
\begin{center}
\includegraphics[scale=.44]{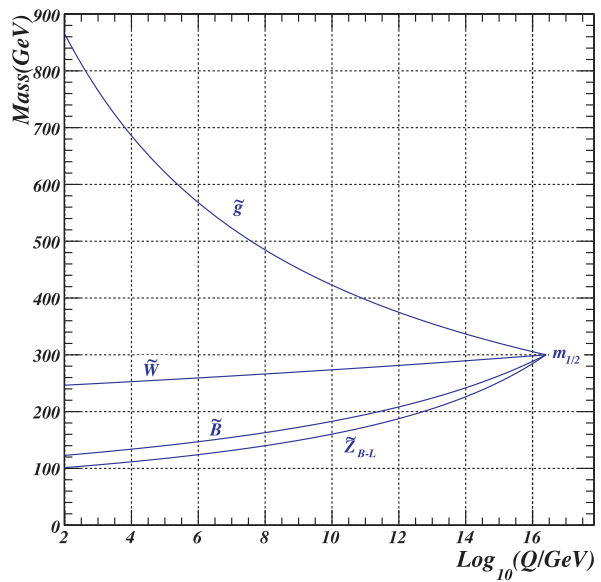}
\caption{Running for the gaugino masses. We have fixed $m_{1/2}(M_{GUT})=300$ GeV in order to reproduce low energy phenomenology.}
\label{fig:gaugino}
\end{center}
\end{figure}

The LHC has started up and it will bring new data for the Higgs searches and also for physics beyond the SM. Supersymmetric particles are one of the most expected signatures that it could be detected in the near future. In order to know the masses of the sparticles at the LHC energies, we need to perform the RGE analysis and solve them down to low energies. 
We start with the gauginos, their one loop $\beta-$functions are given by,
\begin{equation}
 \beta _{M_i}=2c_ig^2_iM_i, \qquad i=1,2,3,B-L.
\end{equation}
We already know the solution for each gauge coupling, so, we need to put the initial condition for, the unified gaugino mass, $m_{1/2}$. To be consistent with the phenomenology at low energy, we impose the condition that the mass of the lightest gaugino should be bigger than 100 GeV. To accomplish  this restriction, we found that $m_{1/2}(M_{GUT})\geq 300$ GeV. In FIG. \ref{fig:gaugino}, we present the running for the gaugino masses with $m_{1/2}(M_{GUT})=300$ GeV. 

The first remarkable fact is that the usual conception of the LSP has been modified. In the MSSM scheme, the $\tilde B$ is considered as the lightest component of the LSP, but in this approach, the lightest component is the corresponding $B-L$ gaugino. Meaning that the contribution to the relic density has to be reconsidered again, because its interactions will contribute to the dark matter fraction in the universe.

On the other hand, in this solution we are producing gluinos of masses higher than 850 GeV. This masses are still in agreement with the recent results presented by CMS and ATLAS experiments at CERN~\cite{cern}.
\begin{figure*}[t!]
\begin{center}
\includegraphics[scale=.44]{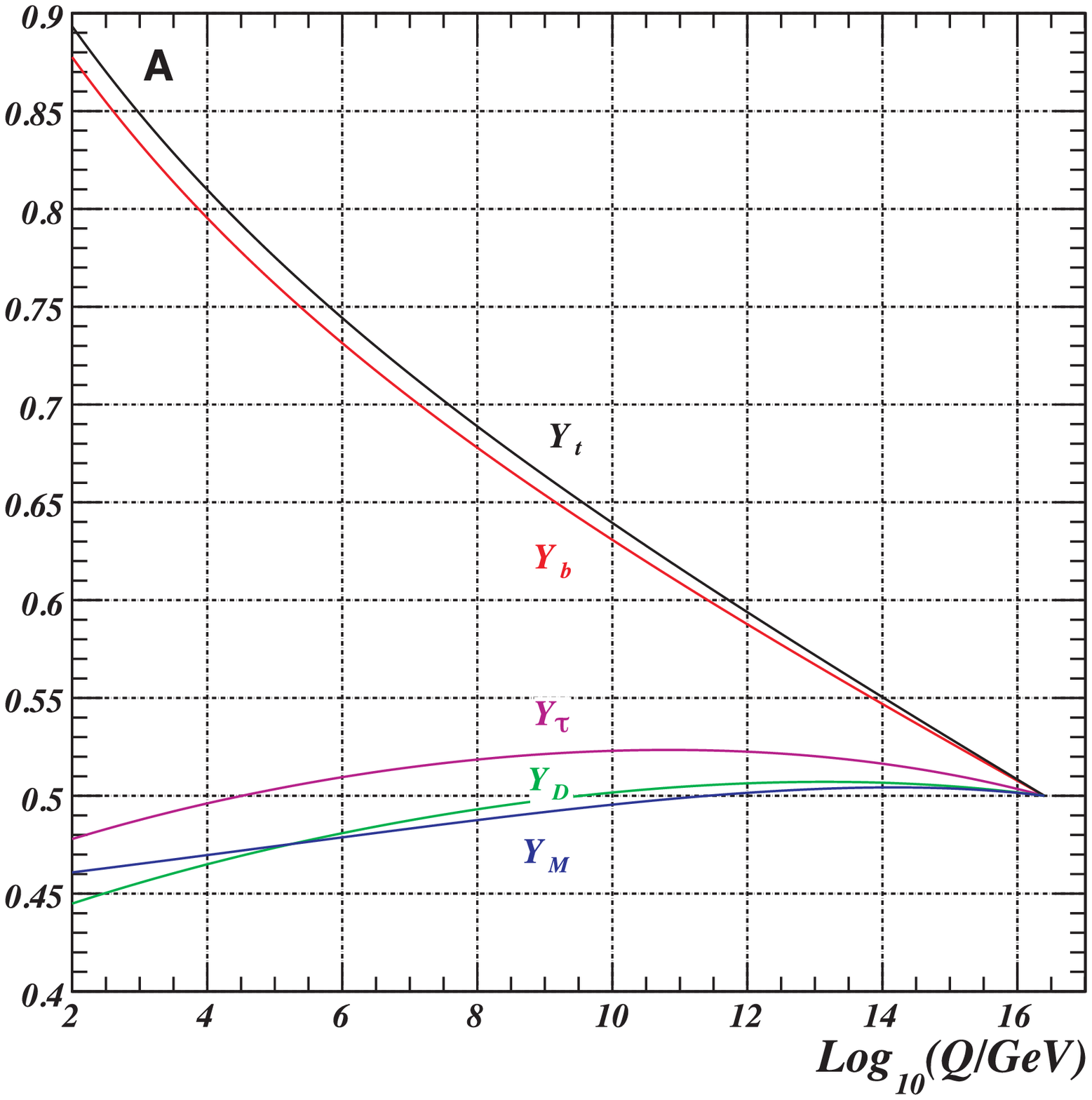}
\includegraphics[scale=.44]{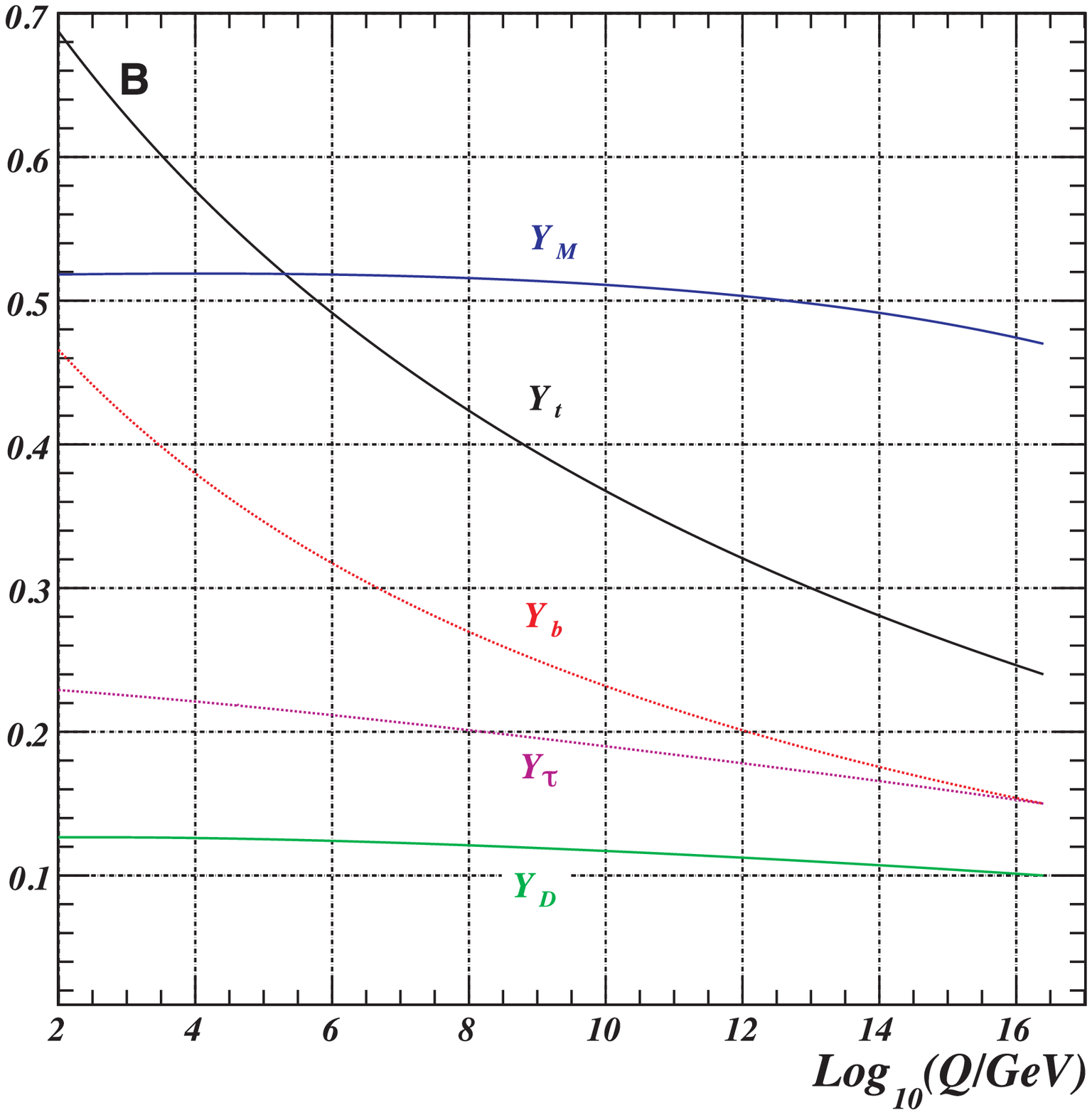}
\includegraphics[scale=.44]{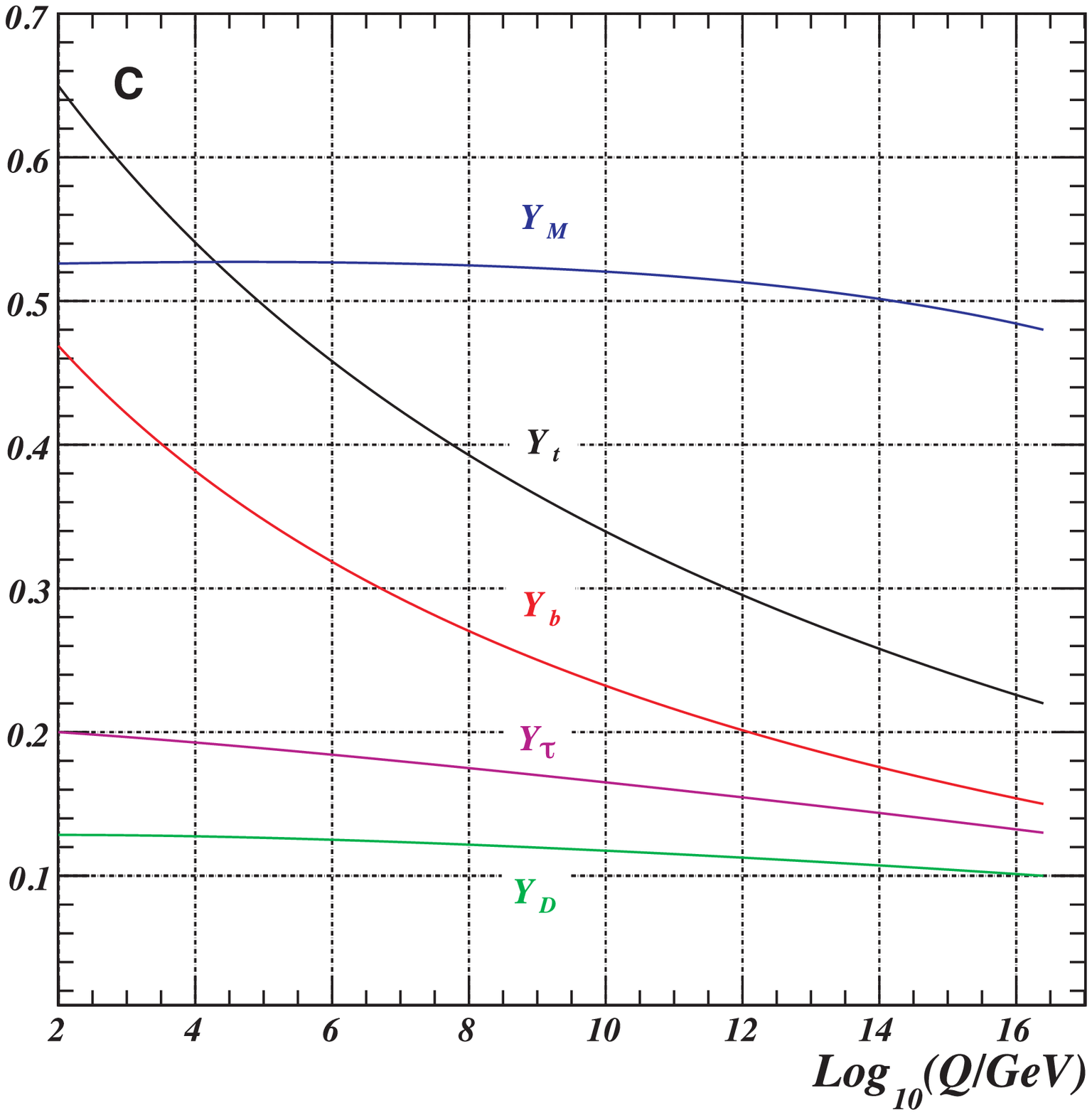}
\caption{Running of the Yukawa's constants in the $\mathcal{G}$-SUSY model for three different scenarios. In the panel A, we plot the running where there is a unification in all of the parameters at $M_{GUT}$ scale. In the panel B, the running is generated by considering the $b-\tau$ unification. Finally, in the panel C we do not consider any unification.}
\end{center}
\end{figure*}

In general, the $\beta-$functions of the mass parameters and the Yukawas are given by,
\begin{eqnarray}
\beta_{y^{ijk}}&=& \gamma^i_n y^{njk} +\gamma^j_n y^{ink}+\gamma^k_n y^{ijn} \\
\beta_{M^{ij}}&=& \gamma^i_n M^{nj} +\gamma^j_n M^{in}
\end{eqnarray}
where the $\gamma^i_j$ are the anomalous dimension matrices associated with the superfields, and at one loop level can be computed using the relation,
\begin{eqnarray}
\gamma^i_j = \frac{1}{16\pi^2}\left[ y^{imn}y_{jmn}^* -2g_a^2 C_a(i) \delta_j^i\right].
\end{eqnarray}
With the previous ingredients we are now able to calculate the $\beta-$functions for the Yukawas and the remaining masses. 

For the Yukawa's evolution, we made the usual approximations,
\begin{eqnarray}
\mbox{\bf Y}_u &\sim& \mbox{diag(0, 0, }y_u), 
\quad \mbox{\bf Y}_d \sim \mbox{diag(0, 0, }y_d), \\
 \mbox{\bf Y}_e &\sim& \mbox{diag(0, 0, }y_e), 
\quad \mbox{\bf Y}_N^D \sim \mbox{diag(0, 0, }y_D), \\
\mbox{\bf Y}_N^M &\sim& \mbox{diag(0, 0, }y_M), 
\end{eqnarray}
and by making these approximations we are just assuming that the third family is the heaviest one
and the main contributions to the running will come only from these families. The contributions to the Yukawa's $\beta-$functions are then expressed as,
\begin{eqnarray}
\Delta \beta_{y_t} &=&  y_t\left\{ y_D^2-\frac{1}{6} g^2_{B-L} \right\}, \\
\Delta \beta_{y_b} &= &-\frac{1}{6}y_b g^2_{B-L} , \\
\Delta \beta_{y_{\tau}} &=& y_{\tau} \left\{ y^2_D -\frac{3}{2}g^2_{B-L}\right\}, 
\end{eqnarray}
and for the Dirac and Mayorana Yukawas, we calculated the complete $\beta-$functions,
\begin{eqnarray}
 \beta_{y_D}&=&y_D\left\{ 4y_D^2+3y^2_t+y_M^2-\frac{3}{5}g_1^2-3g^2_2-\frac{3}{2}g^2_{B-L}\right\},\nonumber\\
\beta_{y_M}&=& y_M^2 \left\{ 3y_M^2+4y^2_D -\frac{9}{2}g^2_{B-L} \right\}. 
\end{eqnarray}

It is already known that the breaking of the electroweak sector is due to the running of top Yukawa and we would like to see if we could get the breaking of $U(1)_{B-L}$ by the contribution of the running of all the Yukawas. Due to the fact that the Yukawa couplings have a direct impact on the masses of the particles, we start from these experimental conditions and run them to high energies. We also can restrict them by imposing a unification or not; the condition that we used is such as the gauge group $\mathcal{G}$ would be embedded in some bigger group ($t-b-\tau$ unification plus a unification of $y_D$ and $y_M$), although we don't know if this is at all possible yet.

By looking FIG. 4, it is clear that even if all the Yukawas begin
in the same point at $M_{GUT}$, $y_{t}$ increase faster than the others, that is expected to generate the electroweak breaking. 
Particularly for $y_D$ and $y_M$, they actually
do not increase, even if in some cases they do it slowly, they cannot reach $y_t$ as we will see. Even if this happens, we 
cannot exclude the possible breaking of $U(1)_{B-L}$ due to the renormalization group equations, because 
the $\beta-$function for the masses receives contributions from all the other sparticles, and from the parameters of the soft breaking
terms which could generate this phenomena.

In FIG. 4, we are considering other two scenarios: {\it a)} If we suppose that at energies beyond $M_{GUT}$, the String Theory is the theory which governs the physics, then we do not need to impose a specific unification to the Yukawa's couplings, so there is none unification of the Yukawas at high energies,  {\it b)} on the other hand, if a Grand Unification Theory is the responsible of the unification of the gauge coupling, we need a group which could permits an embedding such that preserves the Gell-Mann$-$Nishijima formulae
\begin{equation}
Q=I_3+\frac{1}{2}Y.
\end{equation}

In the simplest case, we could suppose that exist a group, $\mathcal{F}$ such that $\mathcal{F}\supset SU(5)\times U(1)_{B-L}\supset \mathcal{G}$. In this example, we have an scenario where $b-\tau$ unification is present at the GUT scale. These other possibilities are plotted in FIG. 4, for comparison. 

The $\mu$-parameters also run over the energy scales. We computed its $\beta-$functions and they are given by,
\begin{eqnarray}
\Delta\beta_{\mu}&=&\mu y_D^2,\\
\beta_{\mu^{\prime}}&=&\frac{1}{2}\mu^{\prime}\left\{y_M^2-3g_{B-L}^2\right\}.
\end{eqnarray}
The $\mu-$parameter is related directly with the phenomenology at low energies with the relation,
\begin{equation}
m^2_Z=\frac{|m^2_{H_d}-m^2_{H_u}|}{\sqrt{1-\sin(2\beta)}}-m^2_{H_u}-m^2_{H_d}-2|\mu|^2, \\ \nonumber 
\end{equation}
where $\tan \beta \equiv \frac{\langle H_{u} \rangle}{\langle H_{d} \rangle}$; that means that $\mu$ should be fixed to give the correct
mass of the $Z$ gauge boson, $m_Z\approx 91$ GeV~\cite{PDG}. For the $\mu '$-parameter holds an analogous relation
but relating the $Z_{B-L}$ gauge boson,
\begin{equation}
M^2_{B-L}=\frac{|m^2_{\sigma_2}-m^2_{\sigma_1}|}{\sqrt{1-\sin(2\beta ')}}-m^2_{\sigma_1}-m^2_{\sigma_2}-2|\mu '|^2,
\end{equation}
where $\tan \beta '\equiv \frac{\langle \sigma_1 \rangle}{\langle \sigma_2 \rangle}$. Although the relation is very simialar, it is more complicated to bound it because
we have no knowledge of the physical mass, but rather only know that $M_{B-L}>60$ GeV, from previous discussion. 

For the parameters in the soft breaking term, we have made the following approximations for the matrices {\bf{h}}$_i$,
\begin{eqnarray}
\mbox{\bf h}_u &\sim& \mbox{diag(0, 0, }a_u), 
\quad \mbox{\bf h}_d \sim \mbox{diag(0, 0, }a_d), \\
 \mbox{\bf h}_e &\sim& \mbox{diag(0, 0, }a_e), 
\quad \mbox{\bf h}_N^D \sim \mbox{diag(0, 0, }a_D), \\
\mbox{\bf h}_N^M &\sim& \mbox{diag(0, 0, }a_M). 
\end{eqnarray}
With these approximations, the contributions due to $U(1)_{B-L}$ to the soft parameters are given by,
\begin{eqnarray}
\Delta\beta_{a_t}&=&  a_t\left\{ y^2_D -\frac{1}{6}g^2_{B-L}\right\} \nonumber\\ 
&&+ y_t\left\{ 2a_Dy_D+\frac{1}{3}g^2_{B-L}M_{B-L} \right\},  \\
\Delta\beta_{a_b}&=&  -\frac{1}{6}a_bg^2_{B-L}
+ \frac{1}{3}y_bg^2_{B-L}M_{B-L},  \\
\Delta\beta_{a_{\tau}}&=& a_{\tau}\left\{ y^2_D -\frac{3}{2}g^2_{B-L}\right\} \nonumber \\
&&+ y_{\tau}\left\{2a_Dy_D+3g^2_{B-L}M_{B-L} \right\}. 
\end{eqnarray}

And, for the extra parameters $a_D$ and $a_M$, the $\beta-$functions become,
\begin{widetext}
\begin{eqnarray}
\beta _{a_D}&=&a_D\left\{ 12y_D^2+3y^2_t+2y^2_M+y^2_{\tau} -\frac{3}{5}g_1^2-3g^2_2-\frac{3}{2}g^2_{B-L}\right\}\nonumber\\
&&+\,\, y_D\left\{6a_ty_t+2a_My_M+a_{\tau}y_{\tau}+ \frac{6}{5} g^2_1M_1+6g^2_2M_2+3g^2_{B-L}M_{B-L} \right\}, \\
\beta _{a_M}&=&a_M\left\{ 15y_M^2+8y^2_D-\frac{9}{2}g^2_{B-L}\right\}+\left\{8a_Dy_D+9g^2_{B-L}M_{B-L} \right\} .
\end{eqnarray}
\end{widetext}

\begin{figure*}
\begin{center}
\includegraphics[scale=.44]{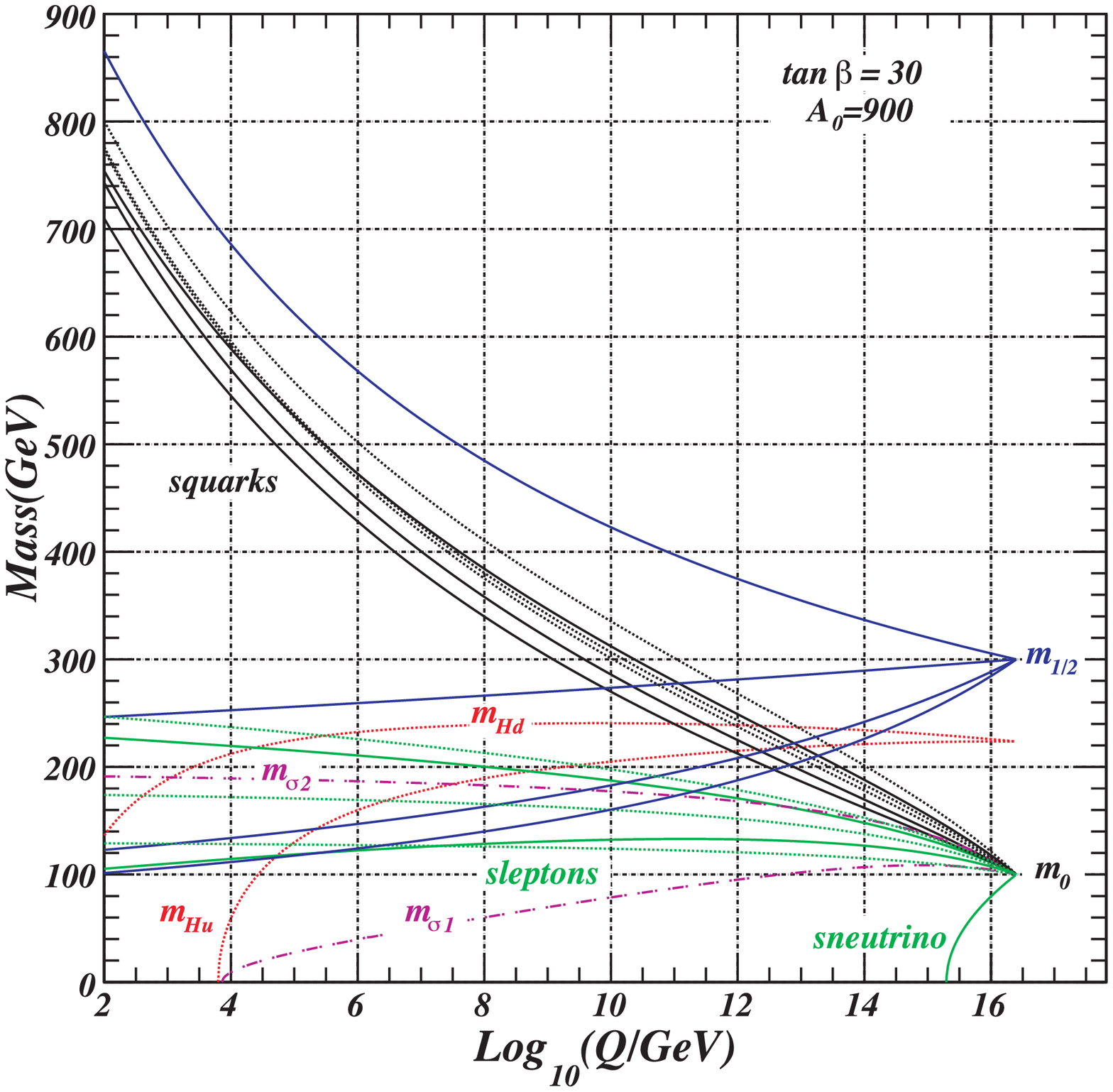}
\includegraphics[scale=.44]{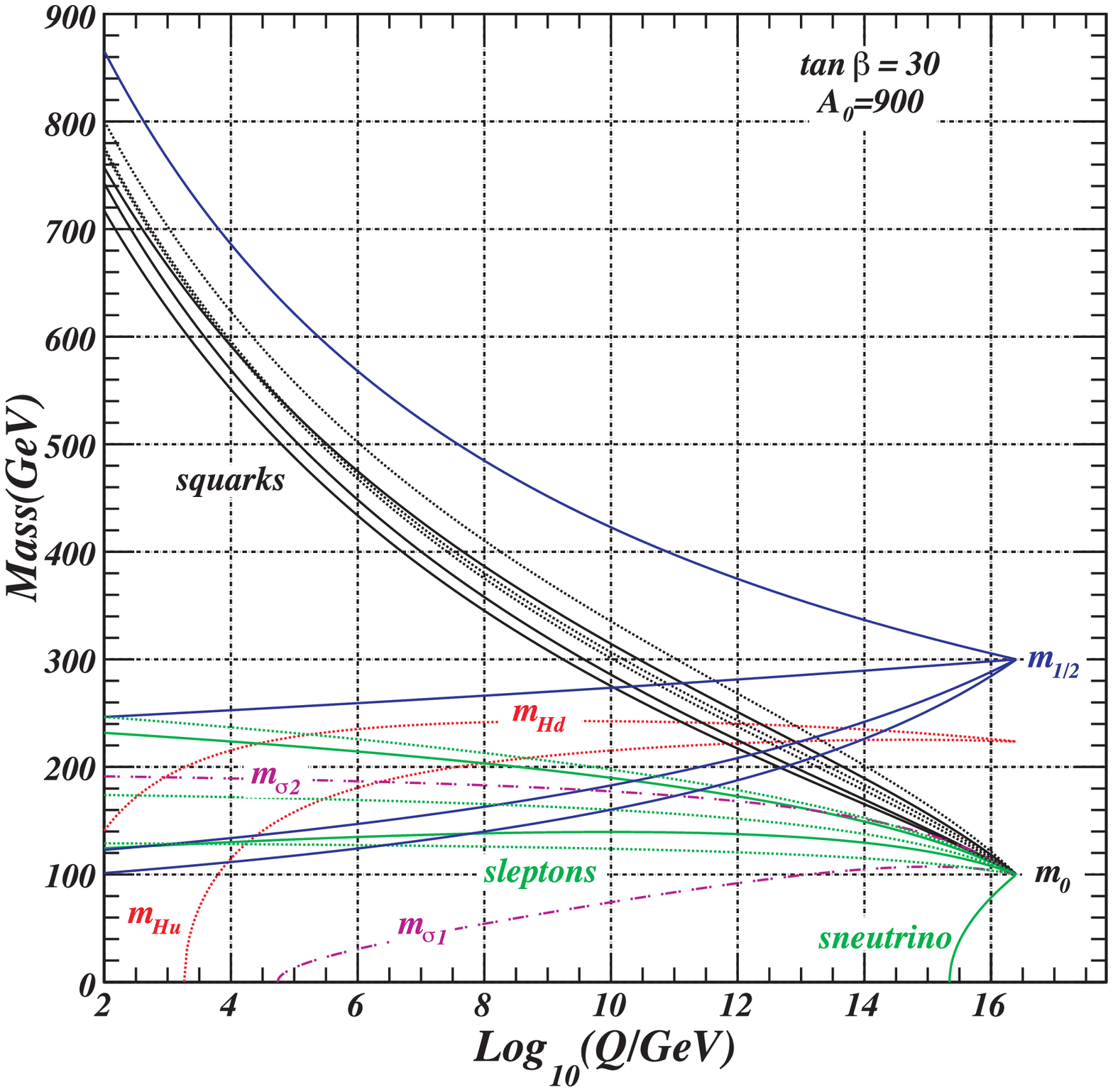}
\caption{Mass spectrum of the sparticles in the $\mathcal{G}$-SUSY model including $b-\tau$ unification (left panel) and without $b-\tau$ unification (right panel).}
\end{center}
\end{figure*}

To be consistent with the previous approximations, we introduce the following anzats for the mass matrices,
\begin{eqnarray}
\mbox{\bf m}_{\bf Q}^{\bf 2} &\sim& \mbox{diag}(m_Q^2 , m_Q^2, m_{Q_3}^2), \\
\mbox{\bf m}_{\bf u}^{\bf 2} &\sim& \mbox{diag}(m_u^2 , m_u^2, m_{u_3}^2), \\
\mbox{\bf m}_{\bf d}^{\bf 2} &\sim& \mbox{diag}(m_d^2 , m_d^2, m_{d_3}^2), \\
\mbox{\bf m}_{\bf L}^{\bf 2} &\sim& \mbox{diag}(m_L^2 , m_L^2, m_{L_3}^2), \\
\mbox{\bf m}_{\bf e}^{\bf 2} &\sim& \mbox{diag}(m_e^2 , m_e^2, m_{e_3}^2), 
\end{eqnarray}
\begin{eqnarray}
\mbox{\bf m}_{\bf N}^{\bf 2} &\sim& \mbox{diag}(m_N^2 , m_N^2, m_{N_3}^2). 
\end{eqnarray}

$\beta-$functions for the scalars are given by,
\begin{eqnarray}
\Delta\beta _{m^2_{H_u}}&=&2y^2_D \left\{ m^2_{H_u} +m^2_{L_3} +m^2_{N_3} \right\} +2a^2_D,\\
\Delta\beta _{m^2_{H_d}}&=&0.
\end{eqnarray}

In order to write the $\beta-$function for the sfermions, we need to define,
\begin{eqnarray}
\mathcal{S}^{\prime}&=&2m^2_{\sigma_2}-2m^2_{\sigma_1}\nonumber\\ &&
+{\mbox Tr}[2\mbox{\bf m}_{\bf Q}^{\bf 2} -2\mbox{\bf m}_{\bf L}^{\bf 2}+\mbox{\bf m}_{\bf u}^{\bf 2}+\mbox{\bf m}_{\bf d}^{\bf 2}-\mbox{\bf m}_{\bf e}^{\bf 2}
-\mbox{\bf m}_{\bf N}^{\bf 2}] \nonumber \\
\end{eqnarray}

Now, the contributions to the sfermions are,
\begin{eqnarray}
\Delta\beta _{m^2_{Q_3}}&=& -\frac{1}{3}g^2_{B-L}M^2_{B-L} +\frac{1}{4} g^2_{B-L}\mathcal{S}^{\prime},\\
\Delta\beta _{m^2_{u_3}}&=&   -\frac{1}{3}g^2_{B-L}M^2_{B-L} +\frac{1}{4} g^2_{B-L}\mathcal{S}^{\prime}, \\
\Delta\beta _{m^2_{d_3}}&=&-\frac{1}{3}g^2_{B-L}M^2_{B-L}+\frac{1}{4} g^2_{B-L}\mathcal{S}^{\prime}, \\
\Delta\beta _{m^2_{e_3}}&=&-3g^2_{B-L}M^2_{B-L} -\frac{3}{4} g^2_{B-L}\mathcal{S}^{\prime},\\
\Delta\beta _{m^2_{L_3}}&=&2y^2_D \left\{ m^2_{H_u} +m^2_{L_3} +m^2_{N_3} \right\} \nonumber\\
&&+2a^2_D-3g^2_{B-L}M^2_{B-L}  -\frac{3}{4} g^2_{B-L}\mathcal{S}^{\prime},
\end{eqnarray}

and, for the new superfields, the $\beta$ functions are given by,
\begin{eqnarray}
\beta _{m^2_{\sigma_1}}&=&2y_M^2\left\{ m^2_{\sigma_1} +m^2_{N_3}  \right\}  \nonumber \\&&
 +2a^2_M -12g^2_{B-L} M^2_{B-L}-\frac{3}{2}g^2_{B-L}\mathcal{S}^{\prime}, \\
\beta _{m^2_{\sigma_2}}&=& -12g^2_{B-L} M^2_{B-L}+\frac{3}{2}g^2_{B-L}\mathcal{S}^{\prime}, \\
\beta _{m^2_{N_3}}&=&4y^2_D\left\{ m^2_{H_u} +m^2_{L_3} +m^2_{N_3} \right\}  \nonumber\\
&& +4y_M^2\{m^2_{\sigma_1}+m^2_{N_3}\} +4(a^2_M+a^2_D) \nonumber\\
&&- 3g^2_{B-L}M^2_{B-L} -\frac{3}{4} g^2_{B-L}\mathcal{S}^{\prime},
\end{eqnarray}

We are choosing the mSUGRA scheme for the running of the parameters; for both scenarios we imposed the following conditions,
\begin{eqnarray}\label{parameters-SUGRA}
&&m_0=100 \,\,\mbox{GeV}, \qquad
m_{1/2}= 300 \,\,\mbox{GeV}, \nonumber \\
&&\tan\beta= 30,  \qquad \qquad
A_0= 900,  \\
&&\mbox{sign } \mu= +.  \nonumber
\end{eqnarray}
We consider the large $\tan\beta$ regime because it is the most common scenario. Nevertheless,  low $\tan\beta$ have to be reanalyzed; High order operators could help to   reach levels above the LEP bounds for the SM-like Higgs as it is pointed out in Ref~\cite{6D}.

In FIG 5, we present the running of the RGE for all the masses of the model.
We can see that the spontaneous breaking of $U(1)_{B-L}$ can be generated in this supersymmetric model, but we also have the right handed neutrino acquiring a vacuum expectation value at higher energies which is due to the running of the Yukawa's constants. Singularly, the early breaking of $U(1)_{B-L}$, which should happened at large scale. 

Between the MSSM and these scenarios there are no quite big differences in the masses of the sparticles. In this scenarios we are producing the masses of the sparticles around 50 GeV higher than those from the MSSM for the same initial conditions.
Interestingly, running of the Yukawa parameters are not so different among them, but their implications in terms of the breaking of $U(1)_{B-L}$ and $SU(2)_L\times U(1)_{Y}$ are very different. 

The matter that the sneutrino mass turns to negative values has been pointed out by several authors~\cite{nuvev}. Nevertheless, the crossing point of its mass parameter is taking place at very high energies in both scenarios, and on top of that, the scale is close the GUT scale. So, this feature deserves to be analyzed closely in order to know what will be the order of magnitude of $\langle \tilde N \rangle$ at low energies. We also need to take care of the $\sigma_1$ field because this field shares information with $\tilde N$ via the soft lagrangian and the $F-$terms, therefore, $\langle \sigma_1 \rangle$ will receive contributions from it and it could be perceptible at low energies.


\section{ $\langle \sigma_1 \rangle$ and $\langle\tilde N \rangle$ at low energies}\label{vev}
In order to compute the vacuum expectation values of $\tilde N$ and $\sigma_1$ at low energies we need to compute the corresponding potential to the sneutrino and the $\sigma_1$ fields together. The contributions are coming from the $F$, $D$ terms and the soft breaking lagrangian,
\begin{eqnarray}
V(\tilde N, \sigma_1)&=& \left(|y_M|^2 +\frac{1}{8}g_{B-L}^2\right)|\tilde N|^4 + m_N^2|\tilde N|^2 \nonumber\\
&& + \frac{1}{8}g_{B-L}^2 |\sigma_1|^4 + \left\{\mu' +m_{\sigma_1}^2 \right\} |\sigma_1|^2 \nonumber\\
&& +4 |y_M|^2 |\tilde N|^2|\sigma_1|^2 +a_M \sigma_1|\tilde N|^2
\end{eqnarray}
where we identified $\tilde N_3\equiv \tilde N$. Now that we know the solution of the RGEs, we minimize the potential and we find its values at low energies. 
\begin{figure}[h]
\begin{center}
\includegraphics[scale=0.44]{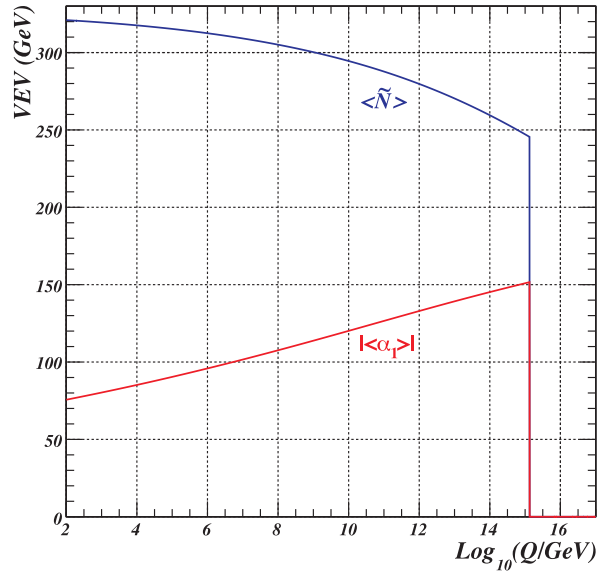}
\caption{General behavior of the sneutrino and the $\sigma_1$ vacuum expectation values. The plot is for the scenario where there is no $b-\tau$ unification.}
\end{center}
\end{figure}

By analyzing FIG. 6, we can see that even if the sneutrino mass becomes negative so fast during the running, the vacuum expectation value does not increase so rapidly and we get a value of the order of 320 GeV at the $m_Z$ scale. On the other hand, the $\langle\sigma_1\rangle$ also starts to have a vev different from zero at the same scale of the $\langle \tilde N\rangle$. Their values are not too high such as we can generate the neutrino masses via two paths, the corresponding to the sneutrino or the corresponding to the $\sigma_1$ field. This two paths will be discussed in a further analysis~\cite{DMsigma}. 
It is remarkable, though, that $\langle \tilde N \rangle$ and $\langle \sigma_1 \rangle$ are generated at such early stage and however, their values remain under control due to SUSY contributions.

\section{Concluding remarks}
In this paper we have studied an extension of the MSSM gauge group, $SU(3)_c\times SU(2)_L \times U(1)_Y\times U(1)_{B-L}$. In order to tackle the neutrino oscillation phenomena, we have added the Dirac and Majorana mass terms into the lagrangian by adding an extra superfield corresponding to right handed neutrinos. The Majorana mass term breaks explicitly the $B-L$ quantum numbers which are preserve in the SM at all orders in the perturbation theory. In order to have an explanation to this breaking, we implemented an analysis based on the RGE. We also computed the associated gauge coupling to $U(1)_{B-L}$ and by imposing the unification at the GUT scale we have found its value at low energies, $g_{B-L}(m_Z)\approx 0.2565$. The possible existence of this gauge group brings a new $Z-$type gauge boson, that might appear at low energy. We have bounded its mass, $M_{B-L} > 60$ GeV, by using the $e^+e^-$ experiment performed by PETRA. 
Running of the sparticles have been perfomed in the mSUGRA scenario in the large $\tan\beta$ regime. The conception that the $SU(2)_L\times U(1)_Y$ provide the lightest supersymmetric particle (LSP) has been modified; by looking at the running of the gaugino masses we could note that the $U(1)_{B-L}$ gaugino is the lightest component of the LSP and it has to be taken into account in order to compute relic density. We have studied in two different schemes for the running of the mass parameters, the $b-\tau$ unification and the no-$b-\tau$ unification. In both cases the breaking of $U(1)_{B-L}$ is reached as it was expected, but due to the different running of the Yukawa's constants, the breaking of the electroweak sector and the $U(1)_{B-L}$ occurs quite differently. In the scenarios studied here, it also happens that the sneutrino acquires a vacuum expectation value at very high energies, close to the GUT scale. Nevertheless, we realized that its running does not grows very rapidly, and its value at low energy remains in a sensitive scale ($\sim320$ GeV); we also calculate $\langle \sigma_1\rangle$ at low energies finding that it also remains in a sizable range ($\sim 70-80$ GeV).
Once we know the order of magnitudes that the parameters can reach in this theory, we can now calculate  the corresponding neutrino masses and the contribution to the dark matter content in the universe due to the extra neutral sparticles.

\begin{acknowledgments}
The authors want to thank to S. Spinner, P. Fileviez-P\'erez and J. de Blas for helpful discussions. R. J. Hern\'andez-Pinto would like to thank E. Camacho-P\'erez, E. A. Garc\'es-Garc\'ia and A. Cervantes-Contreras for their help on making the plots. This work was supported partially by CONACyT, M\'exico, grants 54576 and 132067.

\end{acknowledgments}

\end{document}